\documentclass[prl,twocolumn,showpacs,preprintnumbers,amsmath,amssymb]{revtex4}

\usepackage{graphicx}
\usepackage{dcolumn}
\usepackage{bm}

\begin{document}

\title{Bose-Einstein Condensation of Particle-Hole Pairs in Ultracold Fermionic Atoms Trapped within Optical Lattices}

\author{Chaohong  Lee}
 \altaffiliation{Electronic addresses:
chlee@mpipks-dresden.mpg.de; chleecn@hotmail.com}

\affiliation{Max Planck Institute for the Physics of Complex
Systems, N\"{o}thnitzer Stra$\beta$e 38, D-01187 Dresden, Germany}

\date{\today}

\begin{abstract}

We investigate the Bose-Einstein condensation (BEC, superfluidity)
of particle-hole pairs in ultracold Fermionic atoms with repulsive
interactions and arbitrary polarization, which are trapped within
optical lattices. In the strongly repulsive limit, the dynamics of
particle-hole pairs can be described by a hard-core Bose-Hubbard
model. The insulator - superfluid and charge-density-wave(CDW) -
superfluid phase transitions can be induced by decreasing and
increasing the potential depths with controlling the trapping
laser intensity, respectively. The parameter and polarization
dependence of the critical temperatures for the ordered states
(BEC and/or CDW) is discussed simultaneously.

\end{abstract}

\pacs{03.75.Ss, 32.80.Pj, 71.30.+h}

\maketitle

In recent years, the demonstration of Bose-Einstein condensation
of particle-particle pairs in homogeneous or confined
two-component (spin-1/2) ultracold fermionic atoms has triggered
great theoretical and experimental interest. The BCS-BEC crossover
in ultracold Fermi atomic gases near a Feshbach resonance has been
predicted by using the resonance superfluidity theory
\cite{Holland-Griffin-Carr-Kinnunen}, and been observed in
experiments \cite{Grimm-Jin}. For the fermionic atoms trapped
within optical lattices, the s-wave or d-wave particle-particle
pairs can undergo a phase transition to a superfluid state when
the inter-component interaction is attractive or repulsive
\cite{Lukin-Scalettar}.

With the mechanism of the superfluidity of atom-atom pairs in
ultracold Fermi atomic gases being explored more and more deeply,
the question arises whether the atom-hole pairs in ultracold Fermi
atomic gases can undergo a BEC phase transition similar to
electron-hole pairs \cite{e-h-pair}. Firstly, due to the
essentially non-equilibrium nature of the system of particle-hole
pairs, which usually does not appear in the system of
particle-particle pairs, the particle-hole system becomes an ideal
system for exploring the non-equilibrium quantum mechanics at the
frontier of many-body physics. Additionally, since all condensed
particle-hole pairs can emit photons in tandem, the quantum
coherence in such a condensate will reveal novel optical effects
and nonlinear optical dynamics, which can not be shown by the
condensate of particle-particle pairs. This provides possible
applications in ultrafast digital logical elements and quantum
computation. Furthermore, the controllable interaction strength
and kinetic energy (or hopping strength) in the atomic systems
open up the very exciting potential to investigate macroscopic
quantum coherence and superfluidity under some extreme conditions
that never exist in traditional electronic systems.

In this letter, we show that the atom-hole pairs in arbitrarily
polarized spin-1/2 ultracold Fermi atoms with repulsive
interaction, which are trapped within optical lattices, can
undergo a superfluid phase transition similar to the ultracold
bosonic atoms confined in optical lattices \cite{Zoller-Bloch}. In
the strongly repulsive limit, the dynamics of atom-hole pairs can
be described by a hard-core Bose-Hubbard model. Then, the phase
transition is analyzed with the derived Bose-Hubbard model. At the
same time, the critical temperature for the ordered states
(charge-density-wave and/or Bose-Einstein condensation) is
discussed within the mean-field theory.

Consider an ensemble of arbitrarily polarized ultracold fermionic
atoms occupying two different hyperfine states $|S\rangle$ and
$|P\rangle$, which are trapped within the optical lattices. For
simplicity, we assume the optical lattice potentials as
$V_{0}^{s,p}(\stackrel{\rightharpoonup
}{x})=\sum_{j=1}^{d}V_{0}^{s,p}\cos ^{2}(kx_{j})$. The wave vector
$k$ is determined by the laser wave-lengths, $d$ (=1, 2 or 3) is
the dimension of the optical lattices, and $V_{0}^{s,p}$ are
proportional to the laser intensity. For sufficient low
temperature, all atoms will be localized into the lowest Bloch
band, and the system can be described by an asymmetric
Fermi-Hubbard Hamiltonian \cite{Lukin-Scalettar}
\begin{equation}
\begin{array}{ll}
H= & -\sum\limits_{\left\langle i,j\right\rangle
}(t_{s}f_{si}^{+}f_{sj}+t_{p}f_{pi}^{+}f_{pj})\\
& +\sum\limits_{i}(\epsilon _{s}n_{si}+\epsilon
_{p}n_{pi})+U\sum\limits_{i}n_{si}n_{pi}.
\end{array}
\end{equation}
Here $f_{\sigma i}^{+}$ ($f_{\sigma i}$) are fermionic creation
(annihilation) operators for localized atoms in state
$|\sigma\rangle$ on site $i$, $n_{\sigma i}=f_{\sigma i}^{+}$
$f_{\sigma i}$. The symbol ${\left\langle i,j\right\rangle }$
represents summing over the nearest-neighbors and $\epsilon _{s}$
($\epsilon _{p}$) is the single-atom energy of the atoms in state
$|S\rangle$ ($|P\rangle$). The state-dependent tunneling (hopping)
$t_{s}(t_{p})$ between nearest neighbors can be induced by varying
the potential depth $V_{0}^{s}(V_{0}^{p})$ with controlling the
laser intensity \cite{Lukin-NJP}. Usually, the tunneling strengths
increase with the decrease of potential depths. Below the unitary
limit, the on-site interaction $U$ is proportional to the s-wave
scattering length between atoms occupying different hyperfine
states. The s-wave scattering between atoms occupying the same
hyperfine state is absent due to the Pauli blocking.

The average number of atoms per site (filling number) $n$ and the
polarization $\gamma$ of the considered system are defined as
\begin{equation}
{n=\sum\limits_{i}(n_{si}+n_{pi})/N_{L},}
\end{equation}
\begin{equation}
{\gamma
=\sum\limits_{i}(n_{si}-n_{pi})/\sum\limits_{i}(n_{si}+n_{pi}).}
\end{equation}
The symbol $N_{L}$ is the total number of lattice sites. In the
following, we focus our interests on the half-filled case ($n=1$),
i.e., one atom per site.

The ground state energy per atom depends upon both the
polarization and the energy difference ($\Delta \epsilon =\epsilon
_{p}-\epsilon _{s}$) between two occupied states. With the
definition of polarization, the ground states can be divided into
five different regimes: non-polarized (NP) ground states with
$\gamma=0$, partially polarized in state $|S\rangle$ (PPS) with
$0<\gamma<1$, partially polarized in state $|P\rangle$ (PPP) with
$-1<\gamma<0$, fully polarized in state $|S\rangle$ (FPS) with
$\gamma=1$ and fully polarized in state $|P\rangle$ (FPP) with
$\gamma=-1$. For the one dimensional lattices ($d=1$) with
state-independent hopping ($t_{s}=t_{p}=t$), these regimes can be
exactly obtained with the Bethe-ansatz \cite{Bethe-ansatz}. The
$\Delta \epsilon$ has two critical values
\begin{equation}
\Delta \epsilon _{1}^{c}=\left\{
\begin{array}{lll}
\frac{\left| U\right| }{2}-2t+4t\int_{0}^{\infty
}\frac{J_{1}(w)dw}{w[1+\exp
(\frac{\left| U\right| w}{2t})]} & \text{ for } & U<0, \\
0 & \text{ for } & U>0,
\end{array}
\right.
\end{equation}
and
\begin{equation}
\Delta \epsilon _{2}^{c}=\left\{
\begin{array}{lll}
2t+\left| U\right|  & \text{ for } & U<0, \\
\sqrt{\frac{U^{2}}{4}+4t^{2}}-\frac{U}{2} & \text{ for } & U>0,
\end{array}
\right.
\end{equation}
corresponding to the boundaries between different regimes. Here,
$J_{1}(w)$ is the first kind Bessel function with first order. The
non-polarized, partially polarized and fully polarized regimes
satisfy $\left| \Delta \epsilon \right|\leq\Delta \epsilon
_{1}^{c}$, $\Delta \epsilon _{1}^{c}<\left| \Delta \epsilon
\right|<\Delta \epsilon _{2}^{c}$ and $\left| \Delta \epsilon
\right|\geq\Delta \epsilon _{2}^{c}$ respectively (see Fig. 1).
\begin{figure}[h]
\rotatebox{0}{\resizebox *{8.0cm}{3.5cm} {\includegraphics
{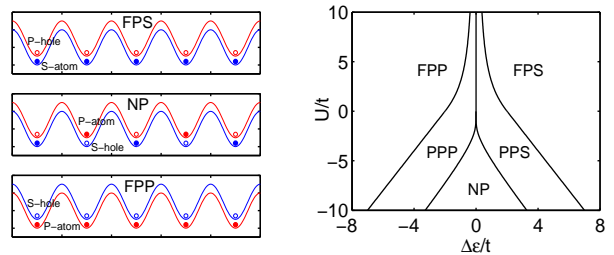}}} \caption{\label{fig:epsart} Left:
Ultracold fermionic atoms in one-dimensional optical lattices with
half-filling and state-independent hopping. The dots and circles
denote the atoms and holes (no atoms) respectively. Right:
Polarization regimes of the ground states for the one-dimensional
lattices with state-independent hopping.}
\end{figure}

In the strongly repulsive limit ($0<t_{s,p}\ll U$), the
Fermi-Hubbard model is equivalent to an effective spin-1/2
Heisenberg model \cite{FHM-HM}. For the case of infinitely
repulsive limit ($U/t_{s,p} \rightarrow + \infty$), the ground
states (lowest energy states) have only one atom for each site,
and their charge degrees of freedom are frozen. Under the strongly
repulsive condition, $U/t_{s,p}\gg 1$, one can introduce the
bosonic operators
\begin{equation}
\begin{array}{l}
b_{j}^{+} \Leftrightarrow f_{sj}^{+}f_{pj},\text{ }b_{j} \Leftrightarrow f_{pj}^{+}f_{sj}, \\
n_{j}=b_{j}^{+}b_{j} \Leftrightarrow
\frac{1}{2}+\frac{1}{2}(n_{sj}-n_{pj}),
\end{array}
\end{equation}
for the atom-hole pairs on site $j$. The operator $b_{j}^{+}
(b_{j})$ creates (annihilates) a pair of S-atom (atom in
$|S\rangle$) and P-hole (hole in $|P\rangle$) on site $j$. These
operators with different lattice indices are commutable. However,
to exclude the multiple occupation at each lattice which comes
from Pauli blocking, the operators with same lattice indices have
a property like Fermi particles. In other words, the interaction
between bosons on the same lattice is infinitely repulsive. Using
the perturbation theory developed by Takahashi \cite{FHM-HM} ($b$
and $b^{+}$ correspond to $\sigma^{-}$ and $\sigma^{+}$), up to
third order terms of the perturbation parameters (hopping
strengths), we obtain the atom-hole pairs obey the hard-core
Bose-Hubbard Hamiltonian
\begin{equation}
{H_{B}= -\mu \sum\limits_{i}n_{i}+J\sum\limits_{\left\langle
i,j\right\rangle }b_{i}^{+}b_{j} + V\sum\limits_{\left\langle
i,j\right\rangle }n_{i}n_{j}.}
\end{equation}
Denoting $t_{p}=\alpha t_{s}=\alpha t$, we obtain the hopping
strength $J=4t_{p} t_{s}/U =4\alpha t^{2}/U $, the nearest
neighbor interaction strength
$V=2(t_{s}^{2}+t_{p}^{2})/U=2(1+\alpha ^{2}) t^{2} /U$ and the
chemical potential $\mu=\Delta \epsilon + ZV/2=\Delta \epsilon +
Z(1+\alpha ^{2}) t^{2} /U$. For the cubic lattices, the total
number of the nearest neighbors $Z$ equals $2d$. The above
hard-core Bose-Hubbard model can be mapped onto an anisotropic
spin-1/2 XXZ Heisenberg model with $J_{xy}=J$, $J_{z}=V$ and an
effective magnetic field $B_{z}=\Delta \epsilon$ \cite{BHM-XXZ}.
The antiferromagnetic-Z order, XY-order and fully magnetized
states in XXZ model correspond to the CDW phase, superfluid phase
and fully polarized insulator phase of the atom-hole pairs,
respectively \cite{BHM-XXZ, Jongh}. The superfluidity means the
Bose condensation of atom-hole pairs in their momentum spaces.

At zero temperature, the ground states for the atom-hole pairs
have three different phases: (i) charge-density-wave phase similar
to a solid phase with zero polarization ($\gamma =0$) corresponds
to the half-filled case of the hard-core Bose-Hubbard model
($\langle b^{+}b \rangle =1/2$), (ii) Bose-Einstein condensation
phase with non-zero superfluid order parameter $\langle b
\rangle$, and (iii) insulator phase with the largest polarization
($| \gamma |=1$) corresponds to the empty ($\langle b^{+}b \rangle
=0$) or the fully-filled ($\langle b^{+}b \rangle =1$) case of the
hard-core Bose-Hubbard model. The difference between superfluid
and insulator phases indicates that it need a non-fully polarized
atomic gases to support the atom-hole BEC. From the equivalence
between the hard-core Bose-Hubbard model and the spin-1/2 XXZ
Heisenberg model, using the path-integral method
\cite{path-integral}, one can obtain that the fully polarized
insulator phase appears when $\left| \Delta \epsilon \right| /U >
Z(t/U)^{2}(1+\alpha )^{2}$, the BEC phase exists if
$(Z/2)(t/U)^{2}\sqrt{(1-\alpha ^{2})^{2}} < \left| \Delta \epsilon
\right| /U < Z(t/U)^{2}(1+\alpha )^{2}$, and the CDW phase emerges
when $\left| \Delta \epsilon \right|
/U<(Z/2)(t/U)^{2}\sqrt{(1-\alpha ^{2})^{2}}$. The separatrix
between CDW phase and BEC phase corresponds to a first order phase
transition. The points on this separatrix means the coexistence of
both phases, they represents the supersolid phase. These
conditions also show the CDW - superfluid and insulator -
superfluid transitions occur at $\left| \Delta \epsilon \right| /U
= (Z/2)(t/U)^{2}\sqrt{(1-\alpha ^{2})^{2}}$ and $\left| \Delta
\epsilon \right| /U = Z(t/U)^{2}(1+\alpha )^{2}$, respectively.

In FIG. 2, we show the phase diagram for lattices of arbitrary
dimensionality with hopping ratio $\alpha=2$. For fixed values of
hopping ratio $\alpha$, energy difference $\Delta \epsilon$ and
on-site repulsive interaction strength $U$, increasing
(decreasing) the hopping strength $t$ will induce an
insulator-superfluid (solid-superfluid) transition. This means
that Bose condensation of the atom-hole pairs exists for mediate
hopping strengths. For larger (smaller) hopping strength, the
ground states fall into the phase of CDW (fully polarized
insulator). In the case of state-independent hopping, $\alpha =1$,
the CDW region becomes a line localized at $\Delta \epsilon =0$.
This is consistent with the results of the antiferromagnetic phase
in Refs. \cite{Lukin-Scalettar}.
\begin{figure}[h]
\rotatebox{0}{\resizebox *{8.0cm}{6.0cm} {\includegraphics
{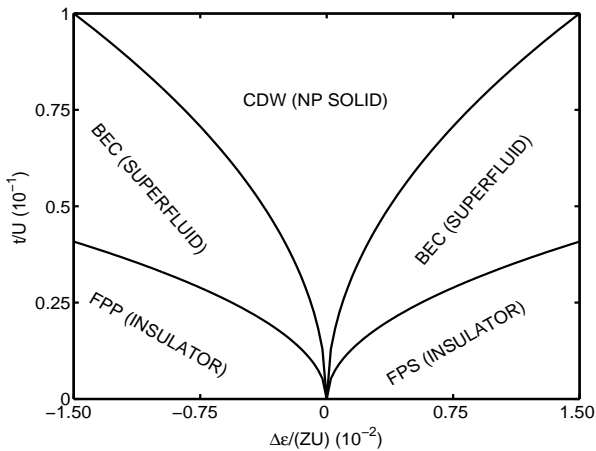}}} \caption{\label{fig:epsart} Zero-temperature
phase diagram of the ground states for the atom-holes in
arbitrarily dimensional lattices with $\alpha=2$.}
\end{figure}

At finite temperatures, due to the thermal fluctuations, the
ordered phases will be destroyed when the temperature is above
some critical temperatures. Within the framework of the mean-field
theory \cite{MFT}, a continuous phase transition between the CDW
and the normal liquid (NL) takes place at the critical temperature
\begin{equation}
T_{CDW}^{C}=\frac{Z}{k_{B}}\cdot \frac{(1+\alpha
^{2})t^{2}}{U}\cdot (1-\gamma ^{2}),
\end{equation}
and a similar phase transition between the superfluid and the
normal liquid occurs at
\begin{equation}
T_{SF}^{C}=\frac{Z}{k_{B}}\cdot \frac{2\alpha t^{2}}{U}\cdot
\frac{\gamma }{arctanh(\gamma )}.
\end{equation}
Here, $k_{B}$ is the Boltzmann constant. The bicritical
polarizations $\gamma=\pm\gamma^{BC}(\alpha) (\gamma^{BC}>0)$ are
given by $T_{SF}^{C}=T_{CDW}^{C}$ and $\mid \gamma \mid \neq 1$.
Below the critical temperatures, there are two coexistence regions
of CDW and superfluid, which correspond to the supersolid regions.
The boundaries between the superfluid and supersolid and between
CDW and supersolid can be obtained by using the Landau expansion
\cite{Landau-expansion}. At zero temperature, the critical
polarization corresponding to the superfluid-supersolid transition
is given as $\gamma^{SFC}=|(1-\alpha)/(1+\alpha)|$. For the case
of state-independent hopping ($\alpha =1)$, the critical
polarization $\gamma^{SFC}=0$, it means that the CDW/solid and
supersolid regions shrink to a line localized at $\gamma=0$. This
indicates that the atom-hole BEC in non-polarized atoms with
state-independent hopping has the highest critical temperature for
the superfluid phase.

The finite temperature phase transitions rely on the hopping ratio
$\alpha$ and the polarization $\gamma$. For state-independent
hopping ($\alpha=1$), CDW - NL and superfluid - NL transitions
occur in non-polarized ($\gamma=0$) and polarized case ($\gamma
\neq 0$), respectively. For state-dependent hopping ($\alpha \neq
1$), the transition routes become more complex. The CDW - NL,
supersolid - CDW - NL, superfluid - supersolid - CDW - NL, and
superfluid -NL phase transitions take place when $\gamma=0$,
$0<|\gamma|\leq \gamma^{SFC}$, $\gamma^{SFC}<|\gamma|<
\gamma^{BC}$, and $|\gamma|\geq \gamma^{BC}$, respectively.

The critical temperatures are determined by both the parameters
and the polarization. The parameter dependence is similar to the
one of Refs. \cite{Lukin-Scalettar}, $T^{C} \propto t^{2}/U$.
Thus, to increase the critical temperatures, one has to decrease
the potential depths $V_{0}^{s,p}$ to increase the hopping
strengths. The polarization dependence of the critical
temperatures and the finite temperature phase diagram are shown in
FIG. 3 for the state-dependent hopping case with $\alpha =2$.
\begin{figure}[h]
\rotatebox{0}{\resizebox *{8.0cm}{6.0cm} {\includegraphics
{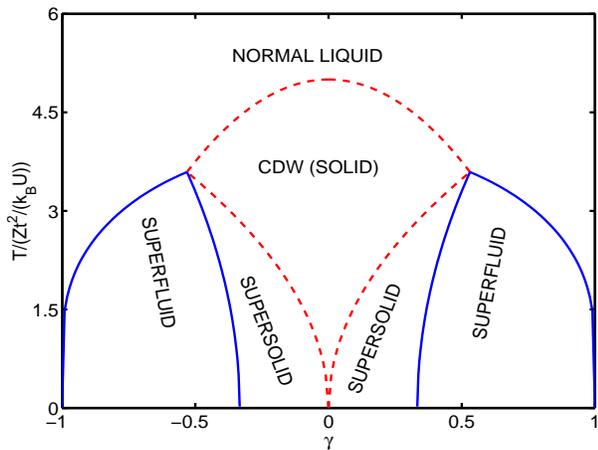}}} \caption{\label{fig:epsart} Mean-field
finite-temperature phase diagram for the atom-hole pairs in
state-dependent hopping case with $\alpha=2$.}
\end{figure}

The previous consideration includes only the terms up to the third
order of the perturbation parameter. Including up to the fifth
order terms, the Hamiltonian (7) reads as
\begin{equation}
\begin{array}{ll}
H_{B}= & -\mu \sum\limits_{i}n_{i}+J_{1}\sum\limits_{\left\langle
i,j\right\rangle }b_{i}^{+}b_{j}+J_{2}\sum\limits_{\left\langle
\left\langle
i,k\right\rangle \right\rangle }b_{i}^{+}b_{k} \\
& + V_{1}\sum\limits_{\left\langle i,j\right\rangle }n_{i}n_{j}+
V_{2}\sum\limits_{\left\langle \left\langle i,k\right\rangle
\right\rangle }n_{i}n_{k}.
\end{array}
\end{equation}
Here, ${\left\langle \left\langle i,k\right\rangle \right\rangle}$
represents summing over the next-nearest-neighbors. The parameters
are determined by $\mu =\Delta \epsilon +Z(V_{1}+V_{2})/2$,
$J_{1}=\frac{4t_{s}t_{p}}{U}[1-\frac{2(t_{s}^{2}+t_{p}^{2})}{U^{2}}]$,
$V_{1}=\frac{4t_{s}t_{p}}{U}(\frac{t_{s}^{2}+t_{p}^{2}}{2t_{s}t_{p}}-
\frac{t_{s}^{4}+t_{p}^{4}+6t_{s}^{2}t_{p}^{2}}{2U^{2}t_{s}t_{p}})$,
$J_{2}=\frac{4t_{s}^{2}t_{p}^{2}}{U^{3}}$, and
$V_{2}=\frac{4t_{s}^{2}t_{p}^{2}}{U^{3}}
(3t_{s}^{4}+3t_{p}^{4}-4t_{s}^{2}t_{p}^{2})/(2t_{s}^{2}t_{p}^{2})$.
The corresponding critical temperatures are formulated as
\begin{equation}
T_{CDW}^{C}=\frac{Z}{2k_{B}}\cdot(V_{1}-V_{2})\cdot(1-\gamma
^{2}),
\end{equation}
\begin{equation}
T_{SF}^{C}=\frac{Z}{2k_{B}}\cdot(J_{1}+J_{2})\cdot\frac{\gamma
}{arctanh(\gamma )}.
\end{equation}
Because $J_{2}\ll J_{1}\approx J$ and $V_{2}\ll V_{1}\approx V$,
the above equations indicate that the critical temperatures are
only shifted a little bit by the high-order terms.

In summary, we have demonstrated the existence of Bose-Einstein
condensation of atom-hole pairs in arbitrarily polarized ultracold
fermionic atoms confined in optical lattices with half filling. In
the strongly repulsive limit, the atom-hole pairs obey a hard-core
Bose-Hubbard Hamiltonian. For a polarized insulator phase, the
particle-hole pairs undergo an insulator - superfluid transition
when the hopping between nearest neighbors is increased. For a
charge-density-wave (CDW) phase, the pairs undergo a CDW -
superfluid transition when the hopping is decreased. The
mean-field results indicate that the finite temperature phase
transition depends upon not only the system parameters but also
the polarization.

To realize the considered model, one can prepare an ultracold
two-component atomic Fermi gases with arbitrary polarization
\cite{Grimm-Jin}, then load them into an optical lattice with one
atom per site. The optical lattices can be produced with a series
of standing-wave lasers. Similar to the experimental realization
of Tonks gas \cite{Tonks-gas} (however the physical details are
only loosely related), the strongly repulsive limit $U/t_{s,p}\gg
1$ can be reached by increasing the s-wave scattering length with
Feshbach resonances \cite{Feshbach-resonance} (but still below the
unitary limit \cite{unitary-limit}) and/or by decreasing the
hopping strengths $t_{s,p}$ with controlling the laser intensity.
The applied magnetic field will also induce a energy difference
between two occupied levels due to the Zeeman effects. To observe
the superfluidity of the atom-hole pairs, one can use Bragg
scattering approach to detect the elementary excitations spectrum
of cold atoms by monitoring the scattered atoms versus the
frequency difference between two lasers \cite{Bragg-scattering},
which form the light grating. In the superfluid phase, it will
appear a distinguished peak corresponding to the collective
Bigoliubov quasi-particle excitations. In a condensed system of
interacting atom-hole pairs within an applied electromagnetic
field, the stimulated two-photon emission (second order harmonics
emission) process can also give a credible evidence for the
atom-hole BEC \cite{JETP}. In this process, the number of the
condensed atom-hole pairs decreases with the emission of the
second order harmonics of the applied light.

The author C. Lee acknowledges useful discussions with professor
Peter Fulde and Dr. Joachim Brand. This work is supported by
foundations of Max Planck Institute for the Physics of Complex
Systems.

\end{document}